# Invisible stimuli, implicit thresholds: Why invisibility judgments cannot be interpreted in isolation

Thomas Schmidt, University of Kaiserslautern, Germany
Faculty of Social Sciences, Experimental Psychology Unit
www.sowi.uni-kl.de/psychologie
thomas.schmidt@sowi.uni-kl.de

Abstract

Some studies of unconscious cognition rely on judgments of participants stating that they have "not seen" the critical stimulus (e.g., in a masked-priming experiment). Trials in which participants gave "not-seen" judgments are then treated as those where the critical stimulus was "subliminal" or "unconscious", as opposed to trials with higher visibility ratings. Sometimes, only these trials are further analyzed, for instance, for unconscious priming effects. Here I argue that this practice requires implicit assumptions about subjective measures of awareness incompatible with basic models of categorization under uncertainty (e.g., modern signal-detection and threshold theories). Most importantly, it ignores the potential effects of response bias. Instead of taking "not-seen" judgments literally, they would better be employed in parametric experiments where stimulus visibility is manipulated systematically, not accidentally. This would allow studying qualitative and double dissociations between measures of awareness and of stimulus processing per se.

**Highlights**

- Examines assumptions underlying "not-seen" judgments (NSJs) in masked priming
- argues that selecting only NSJ trials for further analysis is unwarranted
- concludes that NSJs cannot be interpreted in isolation
- recommends continuous measures of visibility
- such measures should be employed in qualitative and double dissociations

## "Not-seen" judgments and the NSJ-only procedure

Many psychophysical procedures are bewilderingly simple. For instance, when participants are asked to assign numbers to their subjective impression of stimulus magnitude (e.g., calling a very loud tone a "10"), they can do this quite well, and even most reliably when allowed to choose their own number range. Thus, by assembling subjective impressions under numerical categories, valid psychophysical scales of subjective magnitude can be constructed and be used even for seemingly impossible tasks, such as comparing the loudness of a tone to the brightness of a light (Stevens, 1946; Gescheider, 1997). The simplicity of this procedure is deceptive, though. It is based on an intricate theory of psychophysical scaling, involving the mapping of physical properties to sensory responses, sensory responses to subjective magnitudes, and subjective magnitudes to categorization responses, all via various functional relationships (optimistically termed "psychophysical laws", e.g., Weber's, Ekman's, and Stevens'; for an introduction, see Gescheider, 1997).

Hence, subjective judgments of stimulus strength are deeply rooted in psychophysical theory, and they play a vital role in many areas of psychophysics. Recently, however, some studies in the domain of unconscious visual perception make use of subjective classifications in a problematic way. In a first step, their procedure requires participants to perform a rapid response to a target stimulus, followed by a rating of the visibility of a critical stimulus (e.g., a masked prime preceding the target) as either "seen" or "not seen". In a second step, those trials where participants made "not-seen" judgments *(NSJs)* are treated in isolation from those that received any higher visibility ratings, and are interpreted as those trials where the critical stimulus was "invisible" or "subliminal". Nothing is wrong with the first step; it is the second step that invites confusion: taking the "not-seen" judgments literally while underestimating the complexity of the underlying categorization process. The purpose of this paper is to point out that this approach is incompatible with modern psychophysical theorizing.

As a caveat, note that I will not summarily argue against subjective measures of visual awareness (Cheesman & Merikle, 1984) or trial-to-trial measures of visibility (Snodgrass, Bernat, & Shevrin, 2004). Many proposals have been made how a direct measure of visual awareness (e.g., a subjective visibility rating) can be sampled concurrently with an indirect measure of stimulus processing (e.g., a priming effect) on a trial-by-trial basis (see Sandberg, Timmermans, Overgaard, & Cleeremans, 2010, for a comparison of different techniques).[1] For instance, Ramsøy and Overgaard (2004) propose a *perceptual awareness scale (PAS)* where subjective experience of the prime is classified as "no experience", "brief glimpse", "almost clear experience", or "clear experience" on each trial. Similarly, *post-trial wagering* involves a monetary bet on whether or not a stimulus was presented (Persaud, McLeod, & Cowey, 2007) and seems to be highly correlated to PAS ratings and confidence judgments (Sandberg et al., 2010). I have no doubt that trial-to-trial visibility ratings can be informative, provided that the dual-task situation doesn't induce too much interference between direct and indirect tasks. What is of concern here is whether visibility ratings are fine-graded to be later used as a continuous measure of visibility, or whether participants are forced to use a small number of visibility categories, one of which is interpreted in isolation from the others as containing the "unconscious" trials. While my examples all focus on the visual modality, the arguments can of course be generalized to other sensory modalities, memory paradigms, and a range of further research domains.

Following the terminology by Cheesman and Merikle (1984) and Reingold and Merikle (1988) for masked-priming studies, I will call the participant's attempt to identify the critical stimulus or to rate its visibility the *direct task*, as opposed to the *indirect task* that measures whether the critical stimulus has an impact on behavior at all. Direct and indirect tasks give rise to *direct and indirect measures*, for instance, a prime visibility rating and a priming effect in response times. *Objective* direct measures involve judgments of the stimulus; they are called this way because they can be compared with stimulus parameters (for instance, in the *d'* measure of signal detection theory). *Subjective* direct measures involve judgments of perceptual states instead of stimulus parameters.

**Not-seen judgments in practice**

Studies differ greatly in how much information they retain about the subjective stimulus ratings, ranging from fine to coarse categorizations. On one end of the spectrum are studies with very fine-grained, nearly continuous scales. For instance, Sergent and Dehaene (2004) retain all 21 levels of their stimulus visibility ratings, and Scott and Dienes (2008) use a scale of familiarity ratings ranging from 0 to 100. Most studies use considerably fewer rating categories. Zeki and ffytche (1998) employ four categories, including an NSJ category labeled "There was no feeling of something being there. A total guess". Ramsøy and Overgaard (2004) also advocate four categories, with an NSJ category labeled "No experience". Finally, there are studies that use only two categories (e.g., Lau & Passingham, 2006; Ro, 2008; Ro, Singhal, Breitmeyer, & Garcia, 2009), an NSJ and a non-NSJ category.[2] More importantly, studies interpret the NSJ category in radically different ways. Many studies compare some indirect measure of stimulus processing across different categories of the visibility scale to find out how the indirect measure depends on visibility. Such a correlational approach makes full use of all the rating categories in all experimental conditions and is obviously valid. In other studies, however, the NSJ category is interpreted in isolation. Let's look at one of those studies in some detail.

Ro (2008) investigated the role of visual awareness in the redundant target effect. The task was to reach for a single target stimulus (appearing at one of two locations) in the presence or absence of a central distractor stimulus appearing simultaneously and expected to speed responses. Visibility of the distractor was manipulated by a pulse of transcranial magnetic stimulation (TMS) above primary visual cortex. After each reach, participants had "to report whether or not they had perceived the additional central stimulus that was presented in half of the trials" (i.e., a two-category visibility judgment, NSJ or non-NSJ, of the distractor).[3] Ro observed that responses in the presence of the distractor were faster than in its absence, independent of whether TMS was applied. Interestingly, when TMS was applied, the redundant-target effect was the same no matter whether participants indicated that they had or hadn't seen the stimulus. The crucial comparison thus was between three types of trials in the TMS condition: Those where the redundant target was absent, and those where it was present and participants reported to have either "perceived" or "not perceived" the target.

Up to here, everything is fine: the response time effect in the indirect task is basically the same no matter how participants classify the visibility of the critical stimulus in the direct task. At this point, however, the author goes on to interpret that classification in a one-to-one correspondence with perceptual states. First, he defines that "this [...] response was used to sort trials into aware and unaware responses" (p. 380). That this is not mere lab lingo is made clear later: "[...] these stimulus-present, but unconscious trials were phenomenologically identical to the observer as the stimulus-absent trials" (p. 381). This is a far-reaching conclusion involving a whole series of assumptions: 1) that the two rating categories correspond to distinct perceptual states, 2) that one of those states is equivalent to one with no stimulation at all, and 3) that the correspondence is one-to-one in the sense that all responses falling into the same rating category indicate one and the same perceptual state (equivalence classes). But is it really valid to conclude that whenever the NSJ category was chosen the trial looks and feels like one with no stimulus at all?

The problem seems to be that the NSJ category is interpreted not *in relation* to other rating categories, but *in isolation*, as indicating exactly what its label says: "I did not perceive the stimulus". However, what may appear as a mere interpretational imprecision can have grave consequences spilling over into data analysis: In some studies, the authors first record a visibility rating for each trial (or person) and then *discard* all the cases where a non-NSJ rating was given, arguing that in the remainder of cases the critical stimulus was "unconscious".

Two examples of this strategy come from studies on continuous flash suppression. In that paradigm, a masking stimulus is repeatedly flashed to one eye. When a target stimulus is then presented to the other eye, it is measured how long it takes for the target to "break" the flash suppression by the mask and to be consciously perceived (Tsuchiya & Koch, 2005; Feng & He, 2005). Bahrami et al. (2010) used this technique to study the priming of numerosity judgments by different types of primes. In each experiment, several participants were excluded on the basis of visibility ratings and other criteria. Then, in the remaining participants, only those trials were analyzed where participants gave an NSJ rating (exclusion rates ranged from 6 % to 36 % of trials, not counting those from the excluded participants). The stimuli surviving this selection procedure were termed "invisible", and there was no comparison of priming effects across visibility ratings. Feng and He (2005) employed a similar but much exclusive selection procedure for their study of cortical activation by flash-suppressed object pictures. Even though they readily identified rating categories with "visible" and "invisible" stimuli, they provided a full comparison between those stimulus classes.

In some contrast to that, Almeida, Mahon, Nakayama, and Caramazza (2008)

employed continuous flash suppression to study the representation of object categories in the ventral and dorsal visual streams. Targets were pictures of objects or animals. To create target images of low visibility, the authors used a pilot study to select several levels of luminance contrast for the target image. The main task (categorization of the target) was performed on all those contrast levels, and then a separate visibility test was conducted to decide post hoc which of the contrast levels would be selected for analysis. For each participant, the highest contrast level was selected for which the participant's discrimination performance was not different from chance (a rather lenient 61-65 % of correct responses). All contrast levels higher than that were discarded; in an unreported number of cases, entire participants were discarded. It is unclear which proportion of the data ultimately survived to be reported. Again, there was no comparison of priming effects across visibility ratings.

In the following, I will argue that the practice of interpreting one response category in isolation from the others (henceforth called the *NSJ-only procedure*) requires assumptions incompatible with modern psychophysics and doesn't provide a sufficiently solid basis for claims of unconscious cognition. In a next step, I will argue that at the heart of this flawed procedure is a *sampling fallacy*, the erroneous conviction that a restriction of a sample on the basis of measured characteristics is still valid on the population level.

**Threshold theories: From distinct states to distinct responses**

The NSJ-only procedure aims at classifying psychophysical trials by the perceptual states they elicit. Specifically, the procedure tries to tell apart trials where the stimulus was not consciously perceived from those where at least some amount of stimulus awareness was experienced. Because the procedure treats all the trials resulting in NSJs as "unconscious", it obviously assumes some fairly direct mapping between the overt judgment and the covert perceptual state.

In other words, the NSJ-only procedure assumes that the critical stimulus elicits one of two perceptual states (or classes of states), one where the stimulus was not consciously perceived and one where it possibly was. Such a two-state concept of consciousness is the hallmark of a threshold theory. Historically, that brand of theory emerged from detection and discrimination experiments using very weak stimuli at the "absolute threshold" of perception, or very small differences between stimuli at the "discrimination threshold". For instance, *high-threshold theory* assumed that physical stimuli resulted in an internal stimulus representation characterized by a certain strength or magnitude. If the stimulus was strong enough, its internal representation exceeded a detection threshold, passing the observer from an "undetect" state into a "detect" state where he could report the stimulus without making errors, whereas in the undetect state only random guessing was possible. In contrast, the more complex *low-threshold theory* assumed that both states might lead to response errors. Quickly, those simple concepts were elaborated further, taking into account spontaneous variability in the stimulus representation as well as variability in the thresholds. Importantly for our purposes, the development of those models was driven by the need to relate observers' overt responses to their unobservable perceptual states by analyzing the pattern of errors. One consequence of the threshold theories was the development of elaborate guessing models explaining how correct and incorrect responses could occur out of the detect, undetect, or intermediary states (e.g., Atkinson & Kinchla, 1965; Friedman, Carterette, Nakatani, & Ahumada, 1968). Thus, stimulus-absent responses were not taken at face value, but were subject to elaborate correction schemes directly following from the respective theory. Any psychophysical procedure postulating transitions between internal states has to face such complexity.

The NSJ-only procedure requires a threshold model as well, but while the classical models deal with the observer's ability to classify external stimuli (an objective measure), the NSJ-only procedure requires the observer to classify internal states (a subjective measure). An *objective* procedure would compare "stimulus-present" and "stimulus-absent" judgments across trials where the stimulus actually was or wasn't presented. In an objective procedure, the observer may declare a stimulus present when it is in fact absent (a false alarm) or vice versa (a miss), or she may correctly identify a stimulus as present (a hit) or absent (a correct rejection). In a *subjective* procedure involving NSJs, her response may indicate an unconscious state when she is in fact in the conscious state (a false alarm), or vice versa (a miss), or it may correctly indicate the unconscious state (a hit) or conscious state (a correct rejection).

**Signal detection: Categorizing the continuous**

In contrast to threshold theory, signal-detection theory (SDT) does not assume that the internal perceptual states elicited by a stimulus are distinct (Green & Swets, 1966; Macmillan & Creelman, 2005). There are no "detect" and "undetect" states at all. Instead, the theory assumes that stimuli create internal representations that are subject to chance fluctuations (random noise) and that categorical judgments about those internal distributions are based on decision criteria. When attempting to detect a stimulus, the observer tries to decide whether the internal evidence in favor of the stimulus is more likely to come from a statistical distribution of signal plus noise *(S+N)* or from a distribution of noise alone *(N)*. This is done by employing a criterion that would be suited for separating the two overlapping distributions. If the subjective evidence exceeds the criterion, the *S+N* distribution is favored over the *N* distribution because it is more likely to have generated this particular level of subjective evidence for the stimulus.[4] Importantly, SDT recognizes a need for a subsequent step, namely translating the decision process into an appropriate response (e.g., "stimulus present", "stimulus absent"). Under this theory, mistakes occur because trials from the *N* distribution (where no stimulus was presented) sometimes exceed the decision threshold (resulting in false alarms), and because trials from the *S+N* distribution sometimes fail to reach the criterion (resulting in misses).

When employed as an *objective* measure of stimulus detection, SDT is able to distinguish an observer's ability to actually separate the *N* and *S+N* distributions (the sensitivity) from his or her propensity to declare a stimulus present or absent (the response bias). Specifically, bringing observers to adopt a stricter criterion lowers the false-alarm rate at the expense of the hit rate, whereas a more lenient criterion raises both false-alarm and hit rates. When employed as a *subjective* measure of visibility, SDT provides a framework for explaining how stimulus representations of different strengths can be sorted into visibility ratings by means of a set of decision criteria. Here, sensitivity would reflect the observer's ability to accurately categorize his or her perceptual state, and response bias would reflect the overall propensity to choose a specific category.

**Interpreting NSJ trials in isolation**

Let's use the theoretical arsenal of threshold and detection theories to analyze a simple application of the NSJ-only procedure. In principle, the procedure can be applied to experiments where the same weak stimulus is presented again and again, and participants are required to declare it present or absent in each new trial.[5] He has to his disposal a set of response categories that are somehow labeled to reflect an ordered set of states of subjective stimulus visibility, with the lowest rating labeled "I am absolutely certain that no stimulus whatsoever was presented" (our NSJ category). The NSJ-only procedure now proceeds in collecting trials where the NSJ response is given, discarding all those trials with higher visibility ratings, and continuing to analyze the isolated NSJ trials for evidence of indirect effects of stimulus processing (e.g., priming effects).

From both theoretical perspectives, threshold and signal-detection theory, the shortcomings of the NSJ-only procedure become very clear in the case of constant stimulation. Let's start with the threshold account. The procedure implicitly follows a threshold model assuming that the observer was in a distinct perceptual state in all those trials that resulted in NSJs, namely a state where the stimulus was not consciously perceived. Therefore, it faces the very same problems as the classical threshold theories, and it would require the same arsenal of error corrections. However, it is unable to identify the correct threshold model because hits, false alarms, misses, and correct rejections are genuinely unobservable due to the subjective nature of the classification task. Therefore, no guessing correction is applied. Instead, all variance in the responses is directly attributed to a spontaneous flipping back and forth between two perceptual states, "conscious" and "unconscious".

How would signal-detection theory account for response behavior in this kind of experiment? SDT would assume a continuum of perceptual states, resulting in some amount of subjective evidence for the critical stimulus. Response categories would be chosen by means of an ordered set of response criteria. NSJs would be made whenever the evidence for the stimulus fails to exceed a criterion separating the lowermost NSJ category from the next-higher one. If an observer would change that criterion, her likelihood of giving NSJs would increase or decrease, thus being subject to response bias as well as sensitivity. However, because the observer is trying to classify perceptual states instead of external stimuli, the NSJ-only procedure is actually unable to assess the observers' sensitivity, that is, the likelihood that NSJs are really made from the "unconscious" state, as opposed from any different state. Neither does it account for the observer's response bias, that is, the overall propensity for choosing the NSJ category. It doesn't account for the possibility that observers have to choose the NSJ category under uncertainty, that is, under the danger of misclassifying their perceptual state into adjacent response categories (giving an NSJ by false alarm, or failing to give it by a miss). In fact, under invariant stimulation, sensitivity and bias cannot be told apart at all. In this situation, the NSJ-only procedure is running the danger of *capitalizing on chance fluctuations* in either perceptual states or response criteria by declaring any change in responses a change in "visibility".[6]

This situation is essentially unchanged if we assume a variety of stimuli or stimulus conditions. Of course, if different stimuli are employed but presented in blocks, the problems are essentially the same as under constant stimulation. But even if stimuli are randomized across trials, this merely adds another source of variance to the existing structure of different response criteria established for the different response categories. In other words, there will still be false alarms and misses when selecting the NSJ category, but they will arise from variance in the stimuli on top of the variance in perceptual states or response criteria. Of course, the presentation of multiple stimuli is an advantage over constant stimulation because it allows for separation of sensitivity and bias *in stimulus classification*, that is, in an objective measure of stimulus discriminability. But it still doesn't allow for separating those two variables in *perceptual state classification*, that is, the subjective measure of stimulus visibility.

**Multilevel decision models**

I have argued that the NSJ-only procedure is incompatible with basic assumptions of threshold as well as signal detection theory. Those problems are illustrated by two recent theories of NSJ judgments that lead to *multilevel decision models*. I think that both theories support my argument against the unguarded use of NSJ categories because they both point out the difficulties of inferring internal subjective states from overt statements about those states.

For Dienes (2008), what seems like unconscious knowledge is just a discrepancy between decision criteria at different stages: a "first-order", phenomenal stage where the system can be in a set of confidence states, and a "higher-order" stage that represents how the participant classifies ("thinks about") the first-order states. If the higher-order stage has a stricter criterion for declaring the system to be in a "guess" state, there is a gap in classifications where the system "knows something" on one level, but doesn't "know it knows something" on a higher level. For example, Scott and Dienes (2008) investigate the learning of artificial grammars. In each trial, participants have to indicate whether a letter string is grammatical or ungrammatical (with reference to a grammar picked up in the training phase), how familiar the string feels, and the confidence in the grammaticality decision. Confidence is rated in four categories, depending on whether the decision was based on pure guessing (the NSJ category), an intuition about the correct answer, a rule that has been discovered, or memory. The authors found that learner's grammaticality judgments were predicted by familiarity judgments, even when confidence ratings indicated they were guessing. In their model of confidence judgments, Scott and Dienes propose that both confidence judgments and grammaticality judgments are based on the same underlying source of information, a feeling of familiarity. Over the course of an experiment, learners compare familiarity for a new item with respect to the mean of the distribution of familiarity levels encountered so far. If this difference fails to exceed a positive or negative confidence threshold, participants claim to be guessing even though their grammaticality judgments may be better than chance. If the difference exceeds the confidence threshold, then grammaticality judgments are made with increasing accuracy as well as increasing confidence.

This is an interesting approach, but any claim for unconscious learning entirely depends on the authors' "gap argument": the range of familiarity judgments below the confidence threshold that still predicts grammaticality judgments. I wonder whether this gap is an inevitable artifact of comparing a nearly continuous familiarity scale with a coarse, four-category confidence scale. If more confidence ratings were used, it is likely that the gap would become smaller, that is, NSJs would be restricted to smaller familiarity ratings. (At the same time, the correlation between familiarity and grammaticality judgments would tend to disappear because of the decreasing variance of familiarity ratings below the ever-stricter confidence threshold.)

A similar proposal comes from Lau (2008). He argues for a two-level model of signal detection -- the lower level involving the internal signal distributions for stimulus-present and stimulus-absent trials, and the higher level involving *representations* of those internal signal distributions. The idea is that the higher level might misrepresent the lower-level distributions, prompting the observer to use a misleading decision criterion (see Lau, 2008, for an explanation of blindsight as a consequence of a mistaken criterion setting).

The Dienes (2008) and Lau (2008) proposals share the same problem: They put several decision problems in series, one plugged into another, with little hope of identifying process parameters at the lowest level. In Dienes's model, first-order states are freshly partitioned by a new, unknown set of decision criteria. In Lau's model, the type of evidence on which NSJs are based is an unknown function of the first-order subjective evidence. In fact, even though both authors advocate the use of rather coarse subjective measures with NSJ categories, their multilevel decision theories cast additional doubt on the NSJ-only procedure because they basically state that the "true", first-order subjective states are not in a one-to-one correspondence with overt classification behavior.

**The NSJ-only procedure as a sampling fallacy**

The NSJ-only procedure shares many features with a more obvious malpractice in consciousness research, namely to measure identification performance for the critical stimulus in each participant and then to discard those participants who perform above a certain criterion, claiming that for the remainder, the stimulus was unconscious (Schmidt, Haberkamp, & Schmidt, 2011). For example, Cheadle, Parton, Müller, and Usher (2011) claimed that subliminal 50-Hz flicker at one of three spatial positions would enhance target processing at that location by attracting visual attention. In each of three experiments, they discarded those participants that performed above 40% correct target detections so that the remaining group was no longer significantly better than chance, thereby disregarding the data of 4 out of 26, 4 out of 17, 3 out of 14, and 3 out of 11 participants in the various experiments and conditions. (For further discussion of this attentional effect, see Bauer, Cheadle, Parton, Müller, & Usher, 2009; but also van Diepen, Born, Souto, Gauch, & Kerzel, 2010.)

It should be evident that such a procedure capitalizes on chance differences between participants. However, I think there is a deeper misunderstanding involved about the nature of sampled data: Restricting the sample to only the NSJ trials does *not* affect the actual visibility of the stimuli, which is indicated *jointly* by the NSJ and non-NSJ trials. This is why I believe the NSJ-only procedure is based on a sampling fallacy, namely the conviction that a restriction of the sample is somehow still valid at the population level. In fact, selecting only those participants or trials that meet some visibility criterion is analogous to testing a new medication and then discarding all those patients who die from it, concluding that all "suitable" patients do fine under the new drug.

**Inconveniences and absurdities**

Historically, the study of unconscious perception has met with a lot of methodological criticism. Traditionally, unconscious perception is demonstrated by showing that a critical stimulus is below some strict criterion in a direct measure of conscious visibility, while still provoking some nonzero effects in an indirect measure like priming. Critics such as Eriksen (1960), Holender (1986; Holender & Duscherer, 2004), and others have attacked studies using this logic for not presenting any evidence meeting sufficiently strict criteria. For example, in his landmark paper, Marcel (1983) arbitrarily set a discrimination performance of 60 % in the direct measure as a valid criterion for declaring the critical stimulus unconscious (where chance performance would be at 50 %). But is 60 % strict enough? Would 55 % be sufficient, or 51, or 50.1?

One major problem with the NSJ-only procedure is that it can lead to absurd conclusions about unconscious processing even when the critical stimulus is far from invisible by traditional standards. For instance, consider once more a situation where the same weak stimulus is presented on each trial, and assume that the observer uses the NSJ category in 50 % of trials (thus exactly meeting the traditional detection threshold). The proper conclusion would be that the observer is in some specified state of uncertainty about the stimulus, *not* in a state of oblivion. Yet, the NSJ-only procedure would happily continue to analyze the 50 % of "subliminal" trials, ignoring the rest. We can continue this argument by considering smaller and smaller percentages of NSJ judgments. For instance, if the proportion of NSJs would be as low as 10 %, we would certainly conclude that the stimulus is far above the detection threshold, yet the NSJ-only procedure would still invite us to analyze the few NSJ trials for effects of unconscious perception.

The NSJ-only procedure is clearly oblivious to the historical need of the field to proceed with exceptionally stringent criteria of visibility. It also lacks specifics on how it is to be applied. We have just seen that when the percentage of NSJs is low, it is no longer convincing to argue for invisibility in single trials, so that an objective performance criterion may still be necessary on top of subjective judgments -- begging the question, which one? Another open issue is the number of rating categories, because it is quite likely that the percentage of NSJs will depend on the number of categories available (Ramsøy & Overgaard, 2004). Imagine a detection experiment where twenty categories of prime visibility are used, and only the lowest one is the NSJ category. Compare this with an experiment where only the categories "seen" and "not seen" are used. It is likely that the availability of only two rating categories will be an incentive for using the NSJ category more often. Related, how should visibility categories be labeled? It is likely that the specific wording of the NSJ and its adjacent categories will affect the decision criteria separating those categories, especially when the NSJ label assumes the character of a "leading question".

Finally, the NSJ-only procedure is inefficient because it draws on information from only a small percentage of trials. For instance, if stimulus-present and stimulus-absent conditions were intermixed in a detection experiment, the NSJ-only procedure would ignore all of the stimulus-absent trials as well as all those stimulus-present trials where no NSJs were given. Generally, if NSJs are produced in a proportion $r$ of stimulus-present trials occurring with probability $p$, the procedure uses only a proportion $r \cdot p$ of all the available trials. For instance, if there are equal proportions of stimulus-present and stimulus-absent trials, and NSJs are made in 30 % of the stimulus-present trials, then the NSJ-only procedure discards 85 % of the data.

**Alternatives: Qualitative and double dissociations**

Currently, most approaches for demonstrating unconscious cognition aim to show that a direct measure of visual awareness is near chance while an indirect measure of visual processing is above chance. In other words, a stimulus must be shown to be "invisible" (inaudible, unremembered) yet still yield some indirect effect on another measure (e.g., a priming effect in reaction times). This pattern of data constitutes a *simple dissociation* between direct and indirect measures (Schmidt & Vorberg, 2006). The trouble is that from a psychophysical and a statistical point of view, it cannot be proven that a stimulus was invisible -- it can at best be made plausible. Reingold and Merikle (1988) were the first to argue that even when prime identification performance is exactly at chance in a strict objective test, this does not yet imply invisibility of the stimulus unless the measurement process is *exhaustive* with respect to all conscious information in the prime -- that is, the direct measure must be able to detect any change, however small, in conscious awareness. Schmidt and Vorberg (2006) show that an exhaustive measure is a strictly monotonic function of visibility, which means it must be of unlimited sensitivity -- clearly a daunting requirement.[7]

This difficulty persists when subjective measures of visibility are used instead of objective measures based on prime identification performance. Dienes (2008) suggests two criteria for unconscious cognition, both based on subjective measures: a *guessing criterion* which separates NSJs from non-NSJs, and a *zero-correlation criterion*, which consists of demonstrating that participants' confidence is unrelated to their objective performance. Both criteria give valuable information not otherwise obtained from objective measures, but they are both based on a simple-dissociation logic. The guessing criterion needs to be exhaustive in the sense that participants must be accurate in classifying their internal states -- an ability that Dienes (2008) himself doubts in his two-level theory. The zero-correlation criterion is even more problematic because it requires exhaustiveness in *two* measures, the subjective and the objective one.

However, Merikle and Cheesman (1987) outlined a way to circumvent this measurement problem by introducing the concept of *qualitative dissociations.* A qualitative dissociation is a data pattern where an indirect measure of perception (e.g., priming) behaves qualitatively differently on different levels of the direct measure (e.g., visibility). As an example, Merikle and Joordens (1997) used a variant of the Stroop (1935) task where participants responded to the color of red or green target stimuli preceded by the prime words "RED" or "GREEN". The regular Stroop effect features faster responses in consistent trials (where the prime word agrees with the color of the target) than in inconsistent trials. However, when the majority of primes are inconsistent with the target, it is possible that participants adjust their response strategies and become faster in inconsistent than in consistent trials. Interestingly, the authors found this strategic reversal only under conditions where the prime was well visible; when the primes were strongly masked, no reversal occurred. They concluded that processing was qualitatively different under conditions of low and high visibility, respectively.

More generally, if we let awareness vary across experimental conditions, it may be possible to establish a *double dissociation*, which consists in finding an experimental manipulation that changes direct and indirect measures in opposite directions (Schmidt & Vorberg, 2006).[8] For example, a prime-target sequence can lead to response priming effects in reaction times (Klotz & Neumann, 1999; Klotz & Wolff, 1995) that increase with increasing time interval between prime and target onset (Vorberg, Mattler, Heinecke, Schmidt, & Schwarzbach, 2003). However, this increase in priming effects is independent of whether the prime becomes *more* or *less* visible with increasing prime-target interval: The increase in priming is always the same, no matter whether prime visibility is low or high, and no matter whether it is increasing or decreasing with increasing prime-target interval (Vorberg et al., 2003; Mattler, 2003; Albrecht, Klapötke, & Mattler, 2010).[9] It is intuitively clear that if a direct and an indirect measure of prime processing proceed in opposite directions, they cannot both be explained by a single, monotonic source of conscious information about the prime (for mathematical proof, see Schmidt & Vorberg, 2006).[10]

From a measurement-theoretical point of view, double dissociations have many advantages and many surprising features (see Schmidt & Vorberg, 2006, for proofs and details). Firstly, they are obtained by varying the visibility of the prime systematically, not accidentally, in parametric experiments. They do not require and cannot be obtained under conditions of zero visibility of the prime. Secondly, they work under milder measurement assumptions than the traditional zero-awareness criterion; in particular, they require no exhaustiveness assumption. They therefore lead to the surprising conclusion that unconscious stimuli are neither necessary nor desirable for demonstrating unconscious processing.

In another demonstration of a double dissociation, we used a visual lightness illusion (Adelson, 1993) in conjunction with a response priming task (Schmidt, Miksch, Bulganin, Jäger, Lossin, Jochum, & Kohl, 2010). Participants responded to a pair of targets, one light and one dark, by pressing a key on the side of the lighter target. Immediately preceding the targets was a pair of flankers, one light and one dark, with a spatial arrangement consistent or inconsistent with that of the targets. Under control conditions, those flankers induced strong response priming effects, with faster responses in consistent and slower responses in inconsistent trials. Under illusion conditions, however, the physically more luminant flanker could be made to appear either lighter or darker than the physically less luminant flanker. Under such conditions, we found that priming effects always depended on the local luminance contrast of the flankers, not on how the flankers were consciously perceived. In particular, a flanker could *look* lighter than the other flanker, but *prime* as if it was darker (and vice versa). We concluded that rapid motor output was based on a qualitatively different stimulus representation than brightness judgments based on visual awareness.

That study illustrates how double dissociations could go beyond the simple-dissociation paradigm. First, no "invisible" stimuli were needed to dissociate visual awareness from rapid motor activation: The flankers were highly visible and remained on screen unmasked until the participant had responded. Second, the dissociation reveals and is driven by a qualitative difference between conscious and unconscious perception, going beyond a mere existence proof of unconscious processing.

**Conclusions**

The NSJ-only procedure purports to provide a simple solution to the difficult measurement problem of proving a stimulus invisible or "subliminal".[11] But it doesn't deliver -- it just softens the criteria that for skeptics were never strict enough to begin with. Its limitations are revealed by psychophysical standard models, both of the threshold and the signal detection type. My conclusion is that subjective visibility judgments are simply not suited for artificial dichotomies truthfully separating "conscious" from "unconscious" processing. The *prima facie* validity of such dichotomies solely stems from the fact that those category labels bear suggestive names, not because they are theoretically justified. Of course, this doesn't render subjective measures of visibility useless: They can give indispensable information about conscious perception if they are sufficiently fine-grained and, importantly, if single rating categories are never interpreted in isolation from other categories. The selective, isolated use of only a single rating category (or any other subsample of the dataset) is a distortion of the sample.

Instead of taking "not-seen" judgments literally, they would better be employed in parametric experiments where stimulus visibility is manipulated systematically, not accidentally. This would allow studying qualitative and double dissociations between measures of awareness and of stimulus processing per se, going beyond mere existence proofs of unconscious processing and toward discovering the qualitative differences between conscious and unconscious perception.

T. Schmidt: Interpreting invisibility judgments. Cite as: *arXiv:1306.0756v4 [q-bio.NC]*     15

*Brain, 121,* 25-45.

**Author note**

I thank Dirk Kerzel for first alerting me to the problem of NSJ responses and for several rounds of helpful discussions. Thanks also to Guido Hesselmann for pointing me to additional papers using the NSJ procedure in the context of continuous flash suppression. Special thanks also to Dirk Vorberg for critical discussions of this manuscript, and also to Marina Veltkamp and Andreas Weber for helpful comments. Thanks also to my sister Petra Schmidt, who concluded over Indian dinner that the NSJ-only procedure was a "cheat" because it discarded all the data that "didn't fit".

**Footnotes**

[1] The preference for subjective or objective measures is in danger of becoming ideological. For instance, Dienes (2008) strongly advocates the use of subjective measures, claiming that objective measures were promoted by people skeptical about unconscious perception in order to dismiss the hypothesis of unconscious perception more easily. How little is gained from such polemic is illustrated by the author's own introductory example, the blindsight phenomenon, which consists of a dissociation between a subjective and an objective measure. If researchers would exclusively apply subjective measures, blindsight would never have been discovered.

[2] Note that Sergent and Dehaene (2004) argue for a dichotomy of conscious and unconscious states on the basis of their data, but nevertheless try to obtain a fine-grained visibility measure. Importantly, they do not force participants to use a "not-seen" rating category. Ramsøy and Overgaard (2004) have a somewhat incoherent position: They strongly argue against subjective measures involving only two visibility classes, but seem perfectly content with their similarly coarse classification into four visibility classes.

[3] From this description of the participants' instructions, it is not entirely clear what the exact instructions were and whether they entailed any information about the likelihood of stimulus-present and stimulus-absent trials.

[4] Exposition here is for a detection experiment, but it can be easily generalized to a discrimination experiment. Just replace the *N* and *S+N* distributions with $S_1$ and $S_2$ distributions, and change the subjective categorization scale to one with poles of "clearly saw that $S_1$ was presented" and "clearly saw that $S_2$ was presented", with the NSJ category in the center.

[5] This might seem like an odd technique to many, but I have seen it employed in several submitted manuscripts. The method is advocated in a short reply paper by Desender and van den Bussche (2012). Note that studying fluctuations in awareness despite constant stimulation is at the heart of many interesting phenomena, for example, binocular rivalry, motion aftereffects, and the autokinetic effect.

[6] Because the NSJ-only procedure may capitalize on chance fluctuations in stimulus representations, it might miss or even suppress a possible dissociation. When a prime is classified as "unconscious" only because it happens to have a weak stimulus representation, chances are that it will also produce a weak indirect effect in a priming task. So, direct and indirect effects may become correlated because of a confounding variable, representation

strength, which fluctuates across trials and affects both measures.

[7] Reingold and Merikle (1988) state a second criterion, namely that the direct measure must be *exclusive* with respect to conscious information, that is, must not be influenced by any unconscious information. From our formalization, this requirement is not necessary. However, if the *indirect* measure was exclusive with respect to *unconscious* information, this would be a sufficient condition for interpreting a simple dissociation as evidence for unconscious processing (Schmidt & Vorberg, 2006). Personally, I am not optimistic about the exhaustiveness criterion, but argue that in some tasks fast motor measures of stimulus processing might turn out to be exclusive measures of unconscious processing (Schmidt, 2007).

[8] Mathematically, Merikle and Joorden's qualitative dissociation is a special case of Schmidt and Vorberg's double dissociation (see our paper for a proof).

[9] This is true under standard conditions, where only two stimuli are involved and prime visibility is altered by visual backward masking from the subsequent target (e.g., by metacontrast). It is not true when visibility is controlled by degrading the prime stimulus itself (e.g., lowering its energy or contrast), and it gets more complicated when an additional masking stimulus intervenes between prime and target (Lingnau & Vorberg, 2005).

[10] "Monotonic" (precisely, "weakly monotonic") means that in the long run (i.e., the expected values), an increase in the information source cannot lead to a decrease in the attached measure. This assumption is necessary for any sensible measurement process. Dunn and Kirsner (1988) show that without this assumption, double dissociations cannot be interpreted as unequivocal evidence for more than one underlying process. They instead propose an even more intricate dissociation pattern named "reversed association". Under assumptions of weak monotonicity, it too becomes a special case of a double dissociation.

[11] Many authors prefer the term *subliminal* when referring to purportedly unconscious stimuli. That term already implies a threshold model. It also seems to stress the strongly misleading idea that conscious visibility of a stimulus is primarily a function of its physical strength. But some of the most effective methods to render a stimulus invisible do not reduce its energy at all (e.g., backward masking, inattentional blindness). In turn, actually reducing a prime's energy will typically reduce priming as well (Haberkamp, Schmidt, & Schmidt, 2013; Schmidt, Niehaus, & Nagel, 2006; Schmidt & Schmidt, 2013; Vath & Schmidt, 2007; also see Schmidt, Haberkamp, & Schmidt, 2011).